



 \font\twelverm=cmr10 scaled 1200    \font\twelvei=cmmi10 scaled 1200
 \font\twelvesy=cmsy10 scaled 1200   \font\twelveex=cmex10 scaled 1200
 \font\twelvebf=cmbx10 scaled 1200   \font\twelvesl=cmsl10 scaled 1200
 \font\twelvett=cmtt10 scaled 1200   \font\twelveit=cmti10 scaled 1200

 \skewchar\twelvei='177   \skewchar\twelvesy='60


 \def\twelvepoint{\normalbaselineskip=12.4pt
   \abovedisplayskip 12.4pt plus 3pt minus 9pt
   \belowdisplayskip 12.4pt plus 3pt minus 9pt
   \abovedisplayshortskip 0pt plus 3pt
   \belowdisplayshortskip 7.2pt plus 3pt minus 4pt
   \smallskipamount=3.6pt plus1.2pt minus1.2pt
   \medskipamount=7.2pt plus2.4pt minus2.4pt
   \bigskipamount=14.4pt plus4.8pt minus4.8pt
   \def\rm{\fam0\twelverm}          \def\it{\fam\itfam\twelveit}%
   \def\sl{\fam\slfam\twelvesl}     \def\bf{\fam\bffam\twelvebf}%
   \def\mit{\fam 1}                 \def\cal{\fam 2}%
   \def\tt{\twelvett}
   \def\nullspace{\nulldelimiterspace=0pt \mathsurround=0pt }
   \def\big##1{{\hbox{$\left##1\vbox to 10.2pt{}\right.\nullspace$}}}
   \def\Big##1{{\hbox{$\left##1\vbox to 13.8pt{}\right.\nullspace$}}}
   \def\bigg##1{{\hbox{$\left##1\vbox to 17.4pt{}\right.\nullspace$}}}
   \def\Bigg##1{{\hbox{$\left##1\vbox to 21.0pt{}\right.\nullspace$}}}
   \textfont0=\twelverm   \scriptfont0=\tenrm   \scriptscriptfont0=\sevenrm
   \textfont1=\twelvei    \scriptfont1=\teni    \scriptscriptfont1=\seveni
   \textfont2=\twelvesy   \scriptfont2=\tensy   \scriptscriptfont2=\sevensy
   \textfont3=\twelveex   \scriptfont3=\twelveex  \scriptscriptfont3=\twelveex
   \textfont\itfam=\twelveit
   \textfont\slfam=\twelvesl
   \textfont\bffam=\twelvebf \scriptfont\bffam=\tenbf
   \scriptscriptfont\bffam=\sevenbf
   \normalbaselines\rm}


 \def\tenpoint{\normalbaselineskip=12pt
   \abovedisplayskip 12pt plus 3pt minus 9pt
   \belowdisplayskip 12pt plus 3pt minus 9pt
   \abovedisplayshortskip 0pt plus 3pt
   \belowdisplayshortskip 7pt plus 3pt minus 4pt
   \smallskipamount=3pt plus1pt minus1pt
   \medskipamount=6pt plus2pt minus2pt
   \bigskipamount=12pt plus4pt minus4pt
   \def\rm{\fam0\tenrm}          \def\it{\fam\itfam\tenit}%
   \def\sl{\fam\slfam\tensl}     \def\bf{\fam\bffam\tenbf}%
   \def\smc{\tensmc}             \def\mit{\fam 1}%
   \def\cal{\fam 2}%
   \textfont0=\tenrm   \scriptfont0=\sevenrm   \scriptscriptfont0=\fiverm
   \textfont1=\teni    \scriptfont1=\seveni    \scriptscriptfont1=\fivei
   \textfont2=\tensy   \scriptfont2=\sevensy   \scriptscriptfont2=\fivesy
   \textfont3=\tenex   \scriptfont3=\tenex     \scriptscriptfont3=\tenex
   \textfont\itfam=\tenit
   \textfont\slfam=\tensl
   \textfont\bffam=\tenbf \scriptfont\bffam=\sevenbf
   \scriptscriptfont\bffam=\fivebf
   \normalbaselines\rm}


 \def\beginlinemode{\endmode
   \begingroup\parskip=0pt \obeylines\def\\{\par}\def\endmode{\par\endgroup}}
 \def\beginparmode{\endmode
   \begingroup \def\endmode{\par\endgroup}}
 \let\endmode=\par
 {\obeylines\gdef\
 {}}
 \def\singlespace{\baselineskip=\normalbaselineskip}
 
 \def\oneandahalfspace{\baselineskip=\normalbaselineskip
   \multiply\baselineskip by 3 \divide\baselineskip by 2}
 \def\doublespace{\baselineskip=\normalbaselineskip \multiply\baselineskip by
2}
 
 \newcount\firstpageno
 \firstpageno=2

\footline={\ifnum\pageno<\firstpageno{\hfil}\else{\hfil\twelverm\folio\hfil}\fi}

 \let\rawfootnote=\footnote              
 \def\footnote#1#2{{\rm\singlespace\parindent=0pt\rawfootnote{#1}{#2}}}
 \def\raggedcenter{\leftskip=4em plus 12em \rightskip=\leftskip
   \parindent=0pt \parfillskip=0pt \spaceskip=.3333em \xspaceskip=.5em
   \pretolerance=9999 \tolerance=9999
   \hyphenpenalty=9999 \exhyphenpenalty=9999 }
 \def\dateline{\rightline{\ifcase\month\or
   January\or February\or March\or April\or May\or June\or
   July\or August\or September\or October\or November\or December\fi
   \space\number\year}}
 \def\received{\vskip 3pt plus 0.2fill
 \centerline{\sl (Received\space\ifcase\month\or
   January\or February\or March\or April\or May\or June\or
   July\or August\or September\or October\or November\or December\fi
   \qquad, \number\year)}}


 \hsize=6.5truein
 \vsize=8.75truein
 \parskip=\medskipamount
 \twelvepoint            
 \doublespace            
 \overfullrule=0pt       



 \def\title                      
   {\null\vskip 3pt plus 0.2fill
    \beginlinemode \doublespace \raggedcenter \bf}

 \def\author                     
   {\vskip 3pt plus 0.2fill \beginlinemode
    \singlespace \raggedcenter}

 \def\affil                      
   {\vskip 3pt plus 0.1fill \beginlinemode
    \oneandahalfspace \raggedcenter \sl}

 \def\abstract                   
   {\vskip 3pt plus 0.3fill \beginparmode
    \doublespace \narrower ABSTRACT: }

 \def\endtitlepage               
   {\endpage                     
    \body}

 \def\body                       
   {\beginparmode}               

 \def\head#1{                    
   \filbreak\vskip 0.5truein     
   {\immediate\write16{#1}
    \raggedcenter \uppercase{#1}\par}
    \nobreak\vskip 0.25truein\nobreak}

 \def\refto#1{$^{#1}$}           

 \def\references                 
   {\head{References}            

    \beginparmode
    \frenchspacing \parindent=0pt \leftskip=1truecm
    \parskip=8pt plus 3pt \everypar{\hangindent=\parindent}}

 \gdef\refis#1{\indent\hbox to 0pt{\hss#1.~}}    

 \gdef\journal#1, #2, #3, 1#4#5#6{               

     {\sl #1~}{\bf #2}, #3, (1#4#5#6)}           

 \gdef\journ2 #1, #2, #3, 1#4#5#6{               

     {\sl #1~}{\bf #2}: #3, (1#4#5#6)}           

 \def\refstylenp{                
   \gdef\refto##1{ (##1)}                                
   \gdef\refis##1{\indent\hbox to 0pt{\hss##1)~}}        
   \gdef\journal##1, ##2, ##3, ##4 {                     
      {\sl ##1~}{\bf ##2~}(##3) ##4 }}

 \def\refstyleprnp{              
   \gdef\refto##1{ (##1)}                                
   \gdef\refis##1{\indent\hbox to 0pt{\hss##1)~}}        
   \gdef\journal##1, ##2, ##3, 1##4##5##6{               
     {\sl ##1~}{\bf ##2~}(1##4##5##6) ##3}}

 \def\figurecaptions             
   {\endpage
    \beginparmode
    \head{Figure Captions}
 }

 \def\endpage                    
   {\vfill\eject}

 \def\endpaper                   
   {\endmode\vfill\supereject}


 \def\ref#1{Ref. #1}                     
 
 \def\frac#1#2{{\textstyle #1 \over \textstyle #2}}

 \def\sla{\raise.15ex\hbox{$/$}\kern-.57em}
 \def\leaderfill{\leaders\hbox to 1em{\hss.\hss}\hfill}
 \def\twiddle{\lower.9ex\rlap{$\kern-.1em\scriptstyle\sim$}}
 \def\bigtwiddle{\lower1.ex\rlap{$\sim$}}
 \def\gtwid{\mathrel{\raise.3ex\hbox{$>$\kern-.75em\lower1ex\hbox{$\sim$}}}}
 \def\ltwid{\mathrel{\raise.3ex\hbox{$<$\kern-.75em\lower1ex\hbox{$\sim$}}}}
 \def\square{\kern1pt\vbox{\hrule height 1.2pt\hbox{\vrule width 1.2pt\hskip
3pt
    \vbox{\vskip 6pt}\hskip 3pt\vrule width 0.6pt}\hrule height 0.6pt}\kern1pt}

 3

\def\e{\, {\rm e}}
\def\IR{{\rm I}\!{\rm R}}
\def\Gamit{{{\mit\Gamma}^{2n}}}
\def\MG2{{\mit\Gamma}^2}
\def\zbar{{\overline{z}}}
\def\wbar{{\overline{w}}}

\singlespace
\rightline{UBCTP 93-011}
\rightline{November, 1993}
\vskip 0.5 truein
{\let\bf=\bigtenrm
\doublespace
\centerline{\bf Phase Space Isometries and Equivariant}
\centerline{\bf Localization of Path Integrals in Two Dimensions}}
\vskip 0.5truein
\centerline{{\bf Richard J. Szabo} and {\bf Gordon W. Semenoff}}
\vskip 0.2truein
\centerline{\it Department of Physics}
\centerline{\it University of British Columbia}
\centerline{\it Vancouver, British Columbia, V6T 1Z1 Canada}
\vskip 2.0truein
\centerline{\bf Abstract}
\vskip 0.1truein

By considering the most general metric which can occur on a
contractable two dimensional symplectic manifold, we find the most
general Hamiltonians on a two dimensional phase space
to which equivariant localization formulas for
the associated path integrals can be applied.
We show that in the case of a maximally symmetric phase
space the only applicable Hamiltonians are essentially harmonic oscillators,
while for non-homogeneous phase spaces the possibilities are
more numerous but ambiguities in the path integrals
occur. In the latter case we give general formulas for
the Darboux Hamiltonians, as well as the
Hamiltonians which result naturally from a generalized coherent state
formulation of the quantum theory which shows that again the
Hamiltonians so obtained are just generalized versions
of harmonic oscillators. Our analysis and results
describe the quantum geometry of some two dimensional systems.

\vfill\eject

\oneandahalfspace

{\noindent
{\bf 1. Introduction}}
\medskip

There are a number of known examples of quantum systems for which the
Feynman path integral is given exactly by the WKB approximation [1].
This fact has been studied in detail in some recent literature where conditions
under which path integrals could be WKB exact are outlined and general
localization formulas are given [2--9]. This development is particularly
interesting in two respects: It has the possibility of expanding the
number of known examples of quantum systems where the Feynman path
integral can be evaluated exactly and it promises deeper insights into
the geometrical structure of quantum systems. It also forms a convenient
approach to topological quantum field theories [3,4] and
is the basis of a conceptual geometric description of
Poincar\'e supersymmetric quantum field theories [5].

In this Paper, we shall explore the applicability of these recently derived
equivariant localization formulas for quantum mechanical path
integrals by considering the case where the phase space of a quantum
system is a two dimensional contractable symplectic manifold.
Equivariant localization formulas are based on the existence of a metric
on the phase space and also on the requirement that the Hamiltonian must
generate an isometry of this metric. In two dimensions, the
Lie algebra of isometries is either zero, one or three dimensional [10].
We examine the case of a three dimensional isometry group in detail and
show that there are only a small, few-parameter families of Hamiltonians
which fit the localization framework.  We also make some observations
of the case of a one dimensional isometry group wherein the
applications of the localization formulas are the most non-trivial.

The localization formulas are sometimes viewed as infinite
dimensional generalizations of the Duistermaat-Heckman theorem [11,12].
Consider the finite dimensional integral
$$\tilde{Z}(\beta_0)=\int_{\Gamit}{\omega^{\wedge n}\over
n!}\e^{-\beta_0H}\eqno(1.1)$$
over a compact $2n$ dimensional symplectic manifold
$(\Gamit,\omega)$ (i.e. a partition function of classical
statistical mechanics). If the critical points of the
Hamiltonian function $H$ on $\Gamit$ are isolated and the Hamiltonian
vector field $V$
generates a symplectic U(1) group action on $\Gamit$, then the
Duistermaat-Heckman integration formula asserts that (1.1) is given
exactly by a sum over the critical points of $H$:
$$\tilde{Z}(\beta_0)=\sum_{x\in I(H)}{\e^
{-\beta_0 H(x)}\over\beta_0^nW(x)},\eqno(1.2)$$
where $I(H)$ is the critical point set of $H$ and
$$W(x)={{\rm det}^{1/2}_{(\mu\nu)}\left[{\partial^2H\over
\partial x^\mu\partial x^\nu}\right]\over{\rm det}^{1/2}_{(
\mu\nu)}[\omega_{\mu\nu}(x)]}$$
is the determinant arising from a Gaussian integral
near $x$. Here $x^\mu$ are local coordinates on the phase space
$\Gamit$ in which the symplectic 2-form is locally $\omega=
{1\over2}\omega_{\mu\nu}(x)dx^\mu\wedge dx^\nu$,
and the Hamiltonian vector field $V=V^\mu{\partial\over
\partial x^\mu}$ is defined by the equation
$$dH=-i_V(\omega),\eqno(1.3)$$
where $i_V:\Lambda^p(\Gamit)\to\Lambda^{p-1}(\Gamit)$ is the
nilpotent interior multiplication acting on the DeRham complex
$\Lambda^*(\Gamit)$ by contracting forms with the vector
field $V$.

The problem for a path integral which describes the dynamics of a
quantum mechanical system is to evaluate the bosonic phase space
functional integral $$Z(T)=\int_{L\Gamit}{\cal D}x^\mu(t)\,\,{\rm
det}^{1/2}\|\omega_{\mu\nu}
\|\exp\left[i\int_0^Tdt\,\left(\theta_\mu\dot{x}^\mu
-H(x)\right)\right],\eqno(1.4)$$
where $\theta=\theta_\mu(x)dx^\mu$ is the symplectic 1-form which
locally generates $\omega$ as $\omega=d\theta$.
The path integral (1.4) is taken over the
loop space $L{\mit\Gamma}^{2n}$ of the phase space, which is the
space of trajectories $x(t):[0,T]\to{\mit\Gamma}^{2n}$
satisfying periodic boundary conditions $x^\mu(0)=x^\mu(T)$,
and it naturally inherits an infinite dimensional symplectic
structure $\Omega$ from $\Gamit$ defined by lifting of the
symplectic 2-form $\omega$, $\Omega_{\mu\nu}(x;t,t')
=\omega_{\mu\nu}(x(t))\delta(t-t')$.

It has been asserted [2--5,7] that a set of sufficient criteria
for the WKB approximation to the path integral (1.4) to
be exact are

(a) The classical action functional should have only isolated critical
points.

(b) The Hamiltonian vector field should generate a symplectic U(1)
action on the phase space.

(c) The phase space should admit a metric with respect to
which the Hamiltonian vector field is a Killing vector.

{\noindent
There has recently been a generalization of these localization
formulas which avoids the necessity of using action functionals which
have only isolated critical points [6,7], and also for Hamiltonians
which themselves do not satisfy (1.3), but are
generic functionals ${\cal F}(H)$ of some observable $H$
generating the U(1) action on $\Gamit$ as in (1.3) [7]. These more general
integration formulas localize the path integral onto
finite dimensional integrals over submanifolds of the
original phase space $\Gamit$
and can be applied to quantum systems for which the WKB approximation
is unsuitable. Moreover, the same principles can be applied
to Hamiltonians which are constructed from generators of
some non-Abelian Lie algebra acting on ${\mit\Gamma}^{2n}$,
yielding applications such as localization
formulas for two dimensional Yang-Mills theory [9,13] and
representations of the infinitesimal Lefschetz number
of Dirac operators [8].}

It has recently been suggested that for two dimensional
phase spaces, the set of Hamiltonians for which these
criteria apply is rather small [14]. In the following we shall
consider some two dimensional symplectic manifolds which
obey the criterion that there exists at least one Killing
vector field. We first examine the case of two dimensional
maximally symmetric spaces which have three Killing vector fields [10]:
the plane, sphere and Lobaschevsky plane (these are essentially the
contractable spaces of constant curvature). We show
that, if the phase space has the topology of the plane
$\IR^2$, the only ``admissible" Hamiltonian is that of
the displaced harmonic oscillator. In this case, we find
that we could replace the criterion that the Hamiltonian
generates a circle action with the requirement that it
is semi-bounded.

On the sphere we obtain a similar result that the only
Hamiltonian is essentially the height function on the
sphere, a Hamiltonian which is familiar from the coadjoint
orbit quantization of SU(2) and which has been discussed
in connection with the quantization of spin systems [15,16].
It is known that the path integral in this case is given
exactly by the WKB approximation [3,4,17]. The Lobaschevsky plane
turns out to be the same as the sphere except that it
is the coadjoint orbit quantization
of the group SU(1,1) [15--18] \footnote{$^1$}{\tenpoint
Note that homogeneous symplectic manifolds are essentially
coadjoint orbits of Lie groups $G$ [10], and so they can be
expressed group theoretically as coset spaces $G/H$ whose
points can be used to construct appropriate
coherent state representations of $G$ [15,16].}. Here there are two
inequivalent
Hamiltonians, corresponding to a choice of ``spacelike"
and ``timelike" Killing vectors. For these latter two
cases of the sphere and Lobaschevsky plane, the admissible
Hamiltonian in Darboux coordinates is a displaced harmonic
oscillator where, unlike the case of the plane, the
Darboux phase space is not the plane but the unit
disc $D^2=\{z\in{\bf C}^1:z\zbar\leq1\}$ for the sphere
and the complement of the unit disc for the
Lobaschevsky plane.

The two dimensional geometries which have a single Killing
vector are more numerous and we discuss only some
specific examples. We do, however, give a general prescription
for computing the alloted Hamiltonians in Darboux coordinates
and discuss some of the ambiguities associated with the
metric dependence of the localization formulas in these cases which
was first pointed out in [14]. We also consider
a general coherent state formalism [19] corresponding to
the one-parameter isometry group action on the manifold
and show that in this formulation of the quantum theory
the Hamiltonians are again just ``generalized" harmonic oscillators.
This final result includes the three cases mentioned above
as special examples, and is therefore a
general expression for the set of two dimensional Hamiltonian systems
whose phase space path integrals may be equivariantly localized.
Furthermore, we show how these and the more general geometric structures
of the phase space are explicitly realized in the relevant quantized
Hamiltonian systems, which gives a further probe into the
geometrical nature of quantum integrability.

\bigskip
{\noindent
{\bf 2. The Equivariant Localization Principle}}
\medskip

The derivations of the standard localization formulas
are based on equivariant cohomology [12,20] and a
supersymmetry of the underlying Hamiltonian system.
In this Section we begin by briefly sketching how these
ideas lead to the principle of Abelian
equivariant localization for path integrals and the constraints it
imposes on the form of the Hamiltonian
system $(\Gamit,\omega,{\cal F}(H))$. We assume henceforth that
the observable $H$ generates, through the relation (1.3), a
global U(1) action on $\Gamit$.

Let ${\rm Fun}(S^1)$ be the algebra of polynomial
functions on the Lie algebra of U(1) graded so that
an $n$-th order homogeneous polynomial is considered
to be of degree $2n$, and
consider the complex $\Lambda_{\rm inv}^*({\mit\Gamma}^{2n})
\subset\Lambda^*(\Gamit)\otimes\,{\rm Fun}(S^1)$
of differential forms on ${\mit\Gamma}^{2n}$ which are
invariant under the U(1) action on the phase space,
i.e. the equivariant differential forms.
The U(1)-equivariant cohomology of ${\mit\Gamma}^{2n}$
is defined by endowing $\Lambda_{\rm inv}^*({\mit\Gamma}^{2n})$ with
the derivative $D_V=d+i_V$. Then $D_V^2={\cal L}_V$, where
$${\cal L}_V=di_V+i_Vd$$
is the Lie derivative along the Hamiltonian vector field $V$. Thus
$D_V^2=0$ precisely on $\Lambda_{\rm inv}^*({\mit\Gamma}^{2n})$,
and the cohomology of the operator $D_V$ on $\Lambda_{\rm inv}
^*(\Gamit)$ is called the
U(1)-equivariant cohomology of ${\mit\Gamma}^{2n}$,
$H_{\rm inv}^*({\mit\Gamma}^{2n})$. By lifting of the
phase space coordinates we also obtain
the loop space equivariant cohomology $H_{\rm inv}^*
(L\Gamit)$ associated with the loop space Hamiltonian vector
field $V^\mu_L(x;t)=\dot{x}^\mu(t)-V^\mu(x(t))$ corresponding
to the action functional in (1.4).

The action functional
$$S[x]=\int_0^Tdt\,\left(\theta_\mu\dot{x}^\mu-{\cal F}(H)
\right)\eqno(2.1)$$
generates a series $\alpha\in H_{\rm inv}^*(L{\mit
\Gamma}^{2n})$ of equivariantly closed differential forms integrated
over the loop space in the partition function (1.4) [2--7]: $Z(T)=
\int_{L{\mit\Gamma}^{2n}}\alpha$. It follows that for any $s\in\IR^+$ and
for any equivariant form $\beta\in\Lambda_{\rm inv}^*(L{\mit\Gamma}^{2n})$,
the integral
$$\int_{L{\mit\Gamma}^{2n}}\alpha=\int_{L{\mit\Gamma}^{2n}}
\alpha\e^{-sD_{V_L}\beta}\eqno(2.2)$$
is formally independent of the parameter $s$ [9], and so the
right-hand side of (2.2) can be evaluated in the limit $s\to+\infty$
(if this limit in fact exists)
giving a localization of the path integral (1.4) onto the
zeroes of the loop space form $\beta$.

We now come to the main assumption invoked in the equivariant
localization principle. We assume that the phase space $\Gamit$ admits a
globally defined U(1)-invariant Riemannian
structure with metric tensor $g={1\over2}g_{\mu\nu}(x)dx^\mu\otimes dx^\nu$
satisfying
$${\cal L}_Vg=0.\eqno(2.3)$$
In other words, $V$ is a Killing vector of the metric $g$. If
$\Gamit$ is compact and the observable $H$ generates a smooth
global U(1) action on $\Gamit$, then
such a metric can always be obtained from any smooth
metric $g'$ on $\Gamit$ by averaging $g'$ over the group U(1).
In the following we shall always suppose that (2.3) generally holds here,
and the metric $g$ naturally induces a loop space metric
$G_{\mu\nu}(x;t,t')=g_{\mu\nu}(x(t))\delta(t-t')$.

Various localization formulas can now be obtained by
taking $\beta$ to be the dual 1-form $\beta=G(W,\cdot)$ of some loop space
vector field $W$. Then the formula (2.2) becomes
$$\int_{L\Gamit}\alpha=\int_{L\Gamit}\alpha\e^{-sd\beta
-sG(W,V_L)}$$
and yields a localization onto the zeroes of $G(W,V_L)$, which for
suitably chosen $W$ will be the same as the zeroes of $W$.
If the Hessian of $G(W,V_L)$ is non-degenerate
then the large-$s$ limit can be evaluated
by Gaussian integration. For example, if ${\cal F}(H)=H$ and
$W=V_L$ is the loop space Hamiltonian vector field corresponding to the
action functional (2.1), then we can
formally obtain the path integral generalization
of the Duistermaat-Heckman formula (1.2) [2--5]
$$Z(T)=\sum_{x\in I(S)}{\det^{1/2}\Vert\omega_{\mu\nu}\Vert
\over\det^{1/2}\Vert\delta^\mu_{\,\,\nu}\partial_t-\partial_\nu
(\omega^{\mu\lambda}\partial_\lambda H)\Vert}\e^{iS[x]}
\eqno(2.4)$$
where $\omega^{\mu\nu}$ is the matrix inverse of $\omega
_{\mu\nu}$. The formula (2.4) supposes that the determinant
of the Jacobi fields is non-trivial. Thus under these
circumstances the path integral (1.4) localizes onto
classical trajectories and we obtain a WKB localization
of (1.4).

More generally, however, we can set $W^\mu={1\over2}\dot{x}^\mu(t)$ and
formally obtain a localization of (1.4) onto the
time-independent modes $x_0^\mu$ of $x^{\mu}(t)$ [6,7]:
$$Z(T)=\int_{{\mit\Gamma}_0}d\phi_0\,dx_0^\mu\,\,
{\rm det}_{(\mu\nu)}^{1/2}[\omega_{\mu\nu}]
\e^{-iT\left(F(\phi_0)+\phi_0H\right)}\,
{\rm det}_{(\mu\nu)}^{1/2}\left[{{1\over2}(\phi_0\tilde{\Omega}^\mu_{\,\,\nu}
+R^\mu_{\,\,\nu})\over\sinh\left[{T\over2}(\phi_0\tilde{\Omega}^\mu
_{\,\,\nu}+R^\mu_{\,\,\nu})\right]}\right].\eqno(2.5)$$
In (2.5), which is now an ordinary finite dimensional integral
over some submanifold ${\mit\Gamma}_0$
of the phase space $\Gamit$, $F(\phi)$ determines
a functional Fourier transform of the Hamiltonian,
$$\exp\left(-i\int_0^Tdt\,{\cal F}(H)\right)=
\int_{L\Gamit}{\cal D}\phi\,\exp\left(-i\int_0^Tdt\,F(\phi)\right)
\exp\left(-i\int_0^Tdt\,\phi(x)H(x)\right),\eqno(2.6)$$
and $\phi_0$ are the zero modes of the auxilliary field
$\phi(x)$. $R^{\mu}_{\,\,\nu}=R^\mu_{\,\,\nu\lambda\rho}(x)
dx^\lambda\wedge dx^\rho$ is the Riemann curvature
2-form of $g$, while $\tilde{\Omega}^\mu_{\,\,\nu}(x)=2\nabla_\nu
V^\mu(x)$ is the Riemann moment map associated with
the U(1) action on the Riemannian manifold $(\Gamit,g)$ [20]
(Notice that, by the Killing equation (2.3), the matrix
$\tilde{\Omega}$ is antisymmetric and so has only $n(2n-1)$
independent components).

The exponential factor in (2.5) is an equivariant generalization
of the Chern character, while the second determinant is
an equivariant $\hat{A}$-genus [6,7]. Thus in this case we obtain, when
the above formal steps actually carry through, a relatively
simple localization onto equivariant characteristic
classes of the manifold $\Gamit$ (i.e. an equivariant generalization of the
Atiyah-Singer index density for a Dirac operator) [20].
However, although the functional ${\cal F}(H)$ is
{\it a priori} arbitrary, there is no reason to
believe that the functional Fourier transform (2.6)
can exist arbitrarily. If ${\cal F}(H)$ is an
unbounded functional, then a Wick rotation off the real time axis
may produce an analytically continued
propagator which is not a tempered distribution. Thus we expect that the
general localization formula (2.5) is valid only
for Hamiltonians ${\cal F}(H)$ which are bounded functionals of
the observable $H$.

We remark that there are also weaker localization formulas
for Hamiltonian sytems which are not
necessarily completely integrable (i.e. for which
the equivariant localization procedure above
does not carry through). In particular,
consider a Hamiltonian ${\cal F}(H)=H$ with $m<n$
independent integrals of motion $I_k(x)$
in involution,
$$\{H,I_k\}=\{I_k,I_\ell\}=0,\eqno(2.7)$$
where the Poisson bracket is defined by
$$\{f,g\}=-\omega^{-1}(df,dg)\quad;\quad f,g\in\,{\rm C}
^{\infty}(\Gamit).\eqno(2.8)$$
In this case we can take $\beta=\int_0^Tdt\,\dot{I}_k
dI_k$ and formally obtain the integration formula [4]
$$Z(T)=\int_{L\Gamit}{\cal D}x^\mu(t)\,\prod_{k=1}
^m\delta(\dot{I}_k)\e^{iS[x]},\eqno(2.9)$$
a weaker version of the above formulas which localizes
the path integral (1.4) onto the reduced symplectic
subspace of the loop space determined by the
constant values of the conserved charges $I_k$, and it can
be viewed as a quantum generalization of the classical
reduction theorem [21].

The above results indicate that
equivariant cohomology might therefore provide a natural
geometric framework for understanding quantum
integrability (in the sense that the path
integral can be reduced to a finite dimensional
expression). The derivative $D_{V_L}$ above has also
been shown to be a supersymmetry operator on
$\Lambda_{\rm inv}^*(L\Gamit)$ for which the effective
action functional in (1.4) is supersymmetric, with respect to the
Poisson algebra (2.8) given by the underlying
symplectic structure [2--7]. In the next Section we shall
use the global constraint (2.3) on the phase space
geometry to construct the possible Hamiltonians
to which the above localization formulas apply in two dimensions,
and show how the geometry of the phase space is thus explicitly
realized in the underlying Hamiltonian system.

\bigskip

{\noindent
{\bf 3. Equivariant Hamiltonian Systems in Two Dimensions}}
\medskip

We now concentrate on the case of a two dimensional symplectic
manifold $(\MG2,\omega)$, and take $g$ to be of
Euclidean signature. We assume that $\MG2$ is a (compact or
non-compact) Riemann surface which is simply connected, $\pi_1(\MG2)=
H^1(\MG2;{\bf Z})=0$. Since the fundamental group
of $\MG2$ is trivial, it follows from the Riemann uniformization theorem
that via a diffeomorphism and a Weyl transformation
of the coordinates the metric can be put globally into the form
$$g_{\mu\nu}(x)=\e^{\varphi(x)}\delta_{\mu\nu}.\eqno(3.1)$$
Here $\varphi(x)$ is a globally-defined real-valued function on $\MG2$ which
we call the conformal factor of the metric. We put the standard
complex structure on $\MG2$ and define the complex
coordinates $z,\overline{z}=x^1\pm ix^2$, and set
$V^{z,\overline{z}}=V^1\pm iV^2$ and $\partial,\overline{
\partial}={1\over2}(\partial_1\mp i\partial_2)$.

In local coordinates on $\MG2$ the Killing equation
(2.3) has the form
$$g_{\mu\lambda}\partial_\nu V^\lambda+g_{\nu\lambda}
\partial_\mu V^\lambda+V^\lambda\partial_\lambda
g_{\mu\nu}=0\eqno(3.2)$$
which, in complex coordinates for the metric (3.1), becomes
$$\overline{\partial}V^z=0\qquad,\qquad \partial V^{\overline z}
=0\eqno(3.3)$$
$$\partial V^z+\overline{\partial}V^{\overline z}+
V^z\partial\varphi+V^{\overline z}\overline{\partial}
\varphi=0.\eqno(3.4)$$
The equations (3.3) are the Cauchy-Riemann equations
and they say that, in the coordinates
specified by (3.1), the Killing vector field $V^z$ is a
holomorphic function on $\MG2$, while (3.4) is a source
equation for $V^z$ and $V^{\overline z}$.
Rather than attempting to solve the general equation (3.4) directly in the
coordinates involved, we turn to a few facts from the
theory of isometry groups of Riemannian manifolds [10].
For simply connected spaces, the generators
of the isometry group ${\cal I}(\MG2,g)$ of
the phase space form a locally compact Lie algebra
${\cal K}(\MG2,g)=\{U\in T\MG2:{\cal L}_Ug=0\}$ whose
dimension is either 1 or 3 \footnote{$^2$}{\tenpoint
In general, for a Hamiltonian system with $n$ degrees of freedom
we have that $\dim{\cal K}(\Gamit,g)\leq n(2n+1)$.}. We begin by considering
the case $\dim{\cal K}(\MG2,g)=3$ when the phase space
$(\MG2,g)$ is maximally symmetric (i.e. homogeneous
and isotropic about each of its points).

In the case of maximal symmetry it can be shown [10] that
the Gaussian curvature $K$ of $(\MG2,g)$, which is defined
through the Riemann curvature tensor by
$$R_{\lambda\mu\nu\rho}=-K(g_{\lambda\nu}g_{\mu\rho}
-g_{\lambda\rho}g_{\mu\nu}),\eqno(3.5)$$
is constant. Moreover, the
converse is also true, and the constant Gaussian curvature
of the space uniquely determines $(\MG2,g)$, in that any
two maximally symmetric metrics on $\MG2$ with the same $K$
and the same signature are equivalent up to a local
diffeomorphism of $\MG2$. In terms of the conformal factor defined in (3.1)
the Gaussian curvature is
$$K=-{1\over2}\e^{-\varphi}\Delta\varphi,$$
where $\Delta=\partial\overline{\partial}$ is the two dimensional
scalar Laplacian on $\MG2$, and we now consider separately
the cases $K=0$, $K>0$ and $K<0$. Notice that if $\MG2$
were a compact Riemann surface of genus $h$, then an application
of the Gauss-Bonnet theorem would give
$\int_{\MG2}d\,{\rm vol}(g)\,K=4\pi(1-h).$ Thus a maximally symmetric
compact space $\MG2$ of constant negative curvature
must have genus $h\geq2$. Therefore, under our topological assumptions,
in the cases $K=0$ and $K>0$ the phase space $\MG2$
may be either compact or non-compact, while for $K<0$
it is necessarily non-compact.

\medskip
{\noindent
{\it 3.1. Planar Geometries}}

When $K=0$, the space $(\MG2,g)$ is locally flat, and the
conformal factor satisfies the two dimensional Laplace equation
$\Delta\varphi=0$ whose general solutions are
$$\varphi(z,\overline{z})=f_0(z)+\overline{f}_0(\overline{z}),$$
where $f_0(z)$ is an arbitrary holomorphic function on $\MG2$.
Since two dimensional conformal transformations are precisely
the analytic coordinate transformations $z\to w(z)$,
we can conformally map $(\MG2,g)$ onto the plane $\IR^2$ with its
standard flat Euclidean metric $g_{E^2}$. The most general solution to
the Killing equations in these new coordinates ($\varphi=0$ in (3.4))
is then
$$V_0=\left(-i\Omega_0 w+\alpha_0\right){\partial\over
\partial w}+\left(i\Omega_0\wbar+\overline{\alpha}_0\right)
{\partial\over\partial\wbar},\eqno(3.6)$$
where $\Omega_0\in\IR^1$ and $\alpha_0\in{\bf C}^1$
are arbitrary parameters. Notice that the $\Omega_0$-term in (3.6)
generates the single possible rotation
of the plane while the $\alpha_0$-term generates the two possible translations
in Euclidean 2-space $\IR^2$, and they together generate
${\cal I}(\IR^2,g_{E^2})=E^2$, the Euclidean group of the plane.

The local form of the Hamiltonian equations (1.3) is
$$\partial H={i\over2}\omega(w,\overline{w})V^{\overline w}\qquad,
\qquad\overline{\partial}H=-{i\over2}\omega(w,\overline{w})V^w\eqno(3.7)$$
where $\omega={i\over2}\omega(w,\overline{w})dw\wedge
d\overline{w}$. The symplectic 2-form can be explicitly determined by recalling
that the U(1) action on the phase space is required to be symplectic.
This means that $\omega\in\Lambda_{\rm inv}^2(\MG2)$ so that
$$0={\cal L}_V\omega=\partial_\mu\left
(V^\lambda\omega_{\nu\lambda}(x)\right)dx^\mu\wedge
dx^\nu.\eqno(3.8)$$
For the planar
Killing vectors (3.6) the condition (3.8) implies that the function
$\omega(w,\overline{w})$ is constant on $\IR^2$; i.e.
$\omega$ is the volume (or, in this case, the Darboux) 2-form
on flat $\IR^2$.
The equations (3.7) can now be solved on $(\IR^2,g_{E^2})$
by substituting in (3.6) and $\omega(w,\overline{w})=1$, which
gives the observable $H$ on $(\IR^2,g_{E^2})$, and on
the original manifold $(\MG2,g)$ in terms of the metric (3.1),
the most general solution of (1.3) is therefore given by
$$H_0(z,\overline{z})=
\Omega_0w_f(z)\overline{w}_f(\overline{z})+\overline{\alpha}_0w_f(z)
+\alpha_0\overline{w}_f(\overline{z})+C_0,\eqno(3.9)$$
where
$$w_f(z)=\int_{C_z}d\xi\,\e^{f_0(\xi)}\eqno(3.10)$$
is the conformal transformation mapping $(\MG2,g)$ onto flat $\IR^2$
with $C_z\subset\MG2$ a simple curve from some fixed base
point $z_0$ to $z$, and $C_0\in\IR^1$ is a constant of
integration.

\medskip
{\noindent
{\it 3.2. Spherical Geometries}}

Now let us consider the case of a
positive Gaussian curvature $K>0$. In this case
the conformal factor solves the Liouville field equation
$\Delta\varphi(z,\overline{z})=-2K\e^{\varphi(z,\overline{z})}$ [22],
a completely integrable system whose general solutions are
$$\varphi(z,\overline{z})=\log\left[{\partial f_1(z)\overline
{\partial}\,\overline{f}_1(\overline{z})\over\left({K\over4}+
f_1(z)\overline{f}_1(\overline{z})\right)^2}\right]\eqno(3.11)$$
where $f_1(z)$ is any holomorphic function on $\MG2$. By
the essential uniqueness of maximally symmetric spaces, there
exists a local coordinate transformation $(z,\overline{z})\to(w(z,
\overline{z}),\overline{w}(z,\overline{z}))$ mapping $(\MG2,g)$
onto the sphere $S^2$ of radius $K^{-1/2}$ with its
standard round metric $g_{S^2}$ [10].
That $(S^2,g_{S^2})$ is maximally symmetric follows from the fact that it
is the one-point compactification of the maximally symmetric space
$(\IR^2,g_{E^2})$. The Hamiltonian equations (1.3) on
$\MG2$ in this case can therefore be solved by considering the corresponding
problem for a spherical geometry.

The induced metric on $S^2$ from its embedding in Euclidean
3-space $\IR^3$ is
$$g_{S^2}={1\over4K}\left[{\overline{w}^2\over1-w\wbar}dw
\otimes dw+{w^2\over1-w\wbar}d\wbar\otimes d\wbar
+2\left(2+{w\wbar\over1-w\wbar}\right)dw\otimes d\wbar
\right]\eqno(3.12)$$
where $w\wbar\leq1$. By considering the manifest invariances of the
embedding condition $\vec{x}\,^2=K^{-1}$, $\vec{x}\in\IR^3$, simultaneously
with those of the standard Euclidean metric on $\IR^3$, it
is straightforward to show that the Killing vectors of the
metric (3.12) have the general form
$$V_{S^2}=\left(-i\Omega_0w+\alpha_0(1-w\wbar)^{1/2}\right)
{\partial\over\partial w}+\left(i\Omega_0\wbar+\overline
{\alpha}_0(1-w\wbar)^{1/2}\right){\partial\over\partial\wbar}.
\eqno(3.13)$$
The $\Omega_0$-term in (3.13) generates rigid rotations of
the sphere while the $\alpha_0$-term generates the two independent
quasi-translations on $S^2$, and the
isometry group of the sphere is just
${\cal I}(S^2,g_{S^2})=\,{\rm SO}(3)$.

The U(1)-invariance condition (3.8)
in the case at hand then implies again that the symplectic 2-form
$\omega$ is just the volume form on $(S^2,g_{S^2})$, i.e. $\omega(w,\wbar)
=K^{-1}(1-w\wbar)^{-1/2}$, which is a non-trivial
element of $H^2(S^2;{\bf Z})={\bf Z}$. Substituting this
and the Killing vectors (3.13) into (3.7) gives the
function $H$ on $(S^2,g_{S^2})$. We can now use
a generalized stereographic projection from the south pole of
$S^2$,
$$(w(z,\zbar),\wbar(z,\zbar))={4K^{-1/2}
\over1+4K^{-1}f_1(z)\overline{f}_1
(\zbar)}(f_1(z),\overline{f}_1(\zbar)),\eqno(3.14)$$
to map (3.12) back onto the original Riemannian
manifold $(\MG2,g)$ defined by (3.1) and (3.11), and
we find that the most general observable $H$ on $(\MG2,g)$ for $K>0$ is
$$H_1(z,\zbar)={\Omega_0\left({K\over4}-f_1(z)\overline{f}
_1(\zbar)\right)\over{K\over4}+f_1(z)\overline{f}_1(\zbar)}+
{\alpha_0\overline{f}_1(\zbar)+\overline{\alpha}_0f_1(z)
\over{K\over4}+f_1(z) \overline{f}_1(\zbar)}+C_0.\eqno(3.15)$$
The case of the sphere $S^2$ corresponds of course to
the choice of $f_1(z)=(K^{1/2}/2)z$ in (3.15).

\medskip
{\noindent
{\it 3.3. Hyperbolic Geometries}}

The construction for the case $K<0$ is identical to
that in Section 3.2 above, except that now we map onto
the maximally symmetric space $H^2$, the
Lobaschevsky plane of constant negative
curvature, with its standard curved metric $g_{H^2}$ [10].
This space is defined by the embedding of the surface
$\eta_{ij}x^ix^j=K^{-1}$, $\vec{x}=\{x^i\}\in\IR^3$,
in $\IR^3$ with flat Minkowskian metric
$\eta_{ij}=\,{\rm diag}(1,1,-1)$, from which it can be
shown that the Killing vectors of the corresponding
induced metric $g_{H^2}$ on $H^2$ have the general form
$$V_{H^2}=\left(-i\Omega_0w+\alpha_0(1+w\wbar)^{1/2}
\right){\partial\over\partial w}+\left(i\Omega_0\wbar
+\overline{\alpha}_0(1+w\wbar)^{1/2}\right)
{\partial\over\partial\wbar}\eqno(3.16)$$
and generate the isometry group ${\cal I}(H^2,g_{H^2})=\,
{\rm SO}(2,1)$. The rest of the above steps now
carry through analogously for the case at hand, and the final result for
the observable $H$ on $(\MG2,g)$ for $K<0$ is {\it identical}
to (3.15). Notice that now, however,
the functions $f_1(z)$ map $(\MG2,g)$ onto the Poincar\'e
disc of radius $|K|^{1/2}/2$ (whereas in the spherical
case the mapping is onto the entire complex plane with the
usual K\"ahler geometry of $S^2$).

\medskip
{\noindent
{\it 3.4. Geometries with Non-maximal Symmetry}}

The above solutions do not, of course, constitute the
complete solution set of (1.3) for the metric (3.1).
We still need to consider the cases where the
Gaussian curvature of $(\MG2,g)$ is a non-constant
function of the coordinates; i.e. $\dim{\cal K}
(\MG2,g)=1$. The case where $(\MG2,g)$ admits only a one-parameter
group of isometries, or equivalently $(\MG2,g)$ contains
a one dimensional maximally symmetric
subspace, is the richest in its applications
to non-trivial problems.

In this case we can change
coordinates $x\to x'$ on $\MG2$ so that the single isometry generator
$V$ has components $V'^1=a_0$, where $a_0\in\IR^1$ is an
arbitrary constant, and $V'^2=0$ (i.e. the Killing
vector of the flat line $\IR^1$), and so that $g_{12}'
(x')=0$. This can be explicitly
accomplished by introducing two differentiable functions $\chi^1$ and
$\chi^2$ on $\MG2$ such that the Jacobian
${\rm det}_{(\mu\nu)}[\partial_\mu\chi^\nu]$ is
non-trivial and such that $\chi^2(x^1,x^2)$ is the
unique solution of the homogeneous first order linear
partial differential equation
$$V(\chi^2)=V^\mu\partial_\mu\chi^2(x^1,x^2)=0.\eqno(3.17)$$
These yields a set of coordinates $x_0'=\chi$ in which the Killing vector
has components $V_0'^2=0$ and $V_0'^1\neq0$. The desired coordinates
are now obtained by defining $x'^1=a_0\int dx_0'^1/V_0'^1$
and $x'^2=x_0'^2=\chi^2$, and then choosing the characteristic
curves of the coordinate $x'^2=\chi^2$ as found from
(3.17) orthogonal to the paths defined by the
isometry generator $V$ (i.e. choosing the initial data
for the solutions of (3.17) to lie on a non-characteristic
surface) to give $g_{12}'=0$. In the new $x'$-coordinates, the Killing
equation (3.2) implies that $g_{11}'$ and $g_{22}'$
are functions only of $x'^2$, and so the phase
space describes a surface of revolution (e.g. a cylinder
or the ``cigar-shaped" geometries used in black hole theories).

In these new coordinates $\omega$ can be taken to be the Darboux 2-form,
and the Hamiltonian equations (1.3) can be solved as
above to give $H$ in terms of the original
coordinates defined by (3.1) as
$$H(x^1,x^2)=a_0\chi^2(x^1,x^2)\eqno(3.18)$$
where the coordinate transformation function
$\chi^2$, determined from (3.17), is constant along the
integral curves of the Killing vector field $V$. Therefore,
if the general metric (3.1) admits a sole isometry, the
Hamiltonians which result from the equivariant localization constraints are
given by the transformation $x\to x'$ in
(3.18) to coordinates in which the corresponding Killing
vector generates translations in
the coordinate $x'^1$ (which is an explicit
U(1) action on $\MG2$). We will discuss what
Hamiltonian systems can explicitly arise
under these circumstances in the next Section. Notice that
in the case of a one-parameter isometry group the equivariant
condition (3.8) does not uniquely determine the symplectic
2-form $\omega$, and in most cases $\omega$ can even be taken
to be the Darboux 2-form in the original coordinates (3.1).

\bigskip

{\noindent
{\bf 4. Integrable Quantum Systems and Coherent State
Formulations}}
\medskip

In the previous Section we have obtained the explicit
geometric dependence of the Hamiltonian systems which result
from the equivariant localization constraints.
We now consider explicitly what Hamiltonians result from the
above analysis as pertaining to the quantization of some
known systems, and examine the exactness of the localization
formulas (2.4) and (2.5) for the corresponding partition
functions $Z(T|\Omega_0,\alpha_0;\MG2,{\cal F}(H))$ in the
maximally symmetric cases and $Z(T|a_0;\chi,{\cal F}(H))$
in the non-homogeneous cases. We also show that these Hamiltonians
can be written as coherent state matrix elements
associated with the canonical Poisson bracket action (2.8)
of the corresponding isometry groups on the phase space [15,16,19].

In the case where the phase space admits only a
one-parameter isometry group the calculations
can be carried out straightforwardly, but in
the maximally symmetric cases where the phase space isometry
group is three dimensional some care must be taken in
applying the localization formulas. This is because the
equivariant localization principle discussed in Section
2 was based on the assumption that the function $H$
generates a global U(1) action on the phase space. However, as
written above the observables $H$ in general are linear combinations
of functions $H_a$ with $dH_a=-i_{V_a}(\omega)$, where
$V_a$ is the vector field on $\MG2$ which is a
generator of the non-Abelian Lie algebra
${\cal K}(\MG2,g)$. While each of the $H_a$ individually
generate U(1) actions on $\MG2$, the collection of them generate
a Hamiltonian action of the isometry group ${\cal I}(\MG2,g)$
represented in ${\rm C}^\infty(\MG2)$ with
Lie bracket given by the Poisson bracket (2.8). To apply the
above formalism we must therefore consider Hamiltonians
which are functionals of only {\it one} of the $H_a$, corresponding
to an Abelian group action. This is also what one expects
from standard integrability theory [22], which tells us that
an integrable Hamiltonian arises by taking functionals of
action variables $I_k$ which are in involution, as in (2.7).
In the present context, this means that an integrable
Hamiltonian is obtained by considering functionals of
Cartan elements of ${\cal K}(\MG2,g)$ only, which
is equivalent to the requirement that $H$ generate a U(1)
action on the phase space. Hamiltonians which
are bilinear functionals of generators of some non-Abelian
Lie group acting on $\MG2$ can also be considered with
modifications of the equivariant localization principle discussed
in Section 2 [9], but we will consider only the Abelian localization formulas
(2.4) and (2.5) in the following.

\medskip
{\noindent
{\it 4.1. Maximally Symmetric Phase Spaces}}

To determine what (flat space) quantum systems (3.9) describes, we
map onto Darboux coordinates, defined by
$\omega={i\over2}dz\wedge d\zbar$, to see what possible
one dimensional quantum mechanical models can arise
from this observable. This is accomplished
by the conformal transformation $z\to w_f(z)$
defined in (3.10) mapping $(\MG2,g)$ onto
$(\IR^2,g_{E^2})$, which gives (3.9) in the Darboux form
$$H_0^D(z,\zbar)=\Omega_0z\zbar+\overline{\alpha}_0z
+\alpha_0\zbar\quad;\quad z\in{\bf C}^1.\eqno(4.1)$$
Equation (4.1) shows that we can obtain only
the harmonic oscillator ($\alpha_0=0$ in (3.9) and
${\cal F}(H)=H$ in (2.1)), and the free particle ($\alpha_0\in\IR^1$,
$\Omega_0=0$ and ${\cal F}(H)=H^2$).

The localization formula (2.5) was considered in [14]
for the harmonic oscillator (which is already known to be WKB
exact) and shown to give the exact result for its
path integral (as found from the WKB formula (2.4) or
the Schr\"odinger equation)
$$Z\Bigl(T\Bigm|{1\over2},0;\IR^2,H\Bigr)={1\over2\sin{T\over2}}.$$
For the free particle, we find from (2.6) with ${\cal F}(H)=H^2$ that $F(\phi)
=\phi^2$, and $R^\mu_{\,\,\nu}=\tilde{\Omega}^\mu_{\,\,\nu}=0$
in (2.5). The $\phi_0$-integral in (2.5) is thus a trivial Gaussian one,
and we end up with the standard free particle partition
function\footnote{$^3$}{\tenpoint Dykstra {\it et al.} [14] verified
the free particle case as well, when ${\cal F}(H)=H$.
However, as we have shown in Section 3 above, the Hamiltonian vector
field corresponding to $p^2$ cannot generate isometries of
any metric of the form (3.1), which is what
leads to the singular metrics used by Dykstra {\it et al.}.
It is possible to go through a similar analysis as in Section 3
using such singular geometries, and obtain the
same results as in [14].}
$$Z\Bigl(T\Bigm|0,{1\over2};\IR^2,H^2\Bigr)=
\int_{-\infty}^{+\infty}dp\,dq\,\e^{-iTp^2}.$$
The remaining Hamiltonian systems, defined
by (3.9), are merely holomorphic copies of the displaced
harmonic oscillators (4.1) (i.e. $z\to z+\alpha_0$)
defined by the conformal transformation (3.10) of the phase space.

Notice that the three independent terms in (4.1) can also
be written as the matrix elements $L(z,\zbar)=
(z|L|z)/(z|z)$ of the usual Heisenberg-Weyl group generators
$L=a^\dagger a,\,a^\dagger,\,a$, respectively, in the
canonical coherent states [15]
$$|z)=\e^{za^\dagger}|0> { } =\sum_{n=0}^\infty{z^n\over\sqrt{n!}}|n>
\quad;\quad z\in{\bf C}^1.$$
The Heisenberg-Weyl algebra is explicitly realized by the
corresponding planar Poisson algebra of the functions
$L(z,\zbar)$ on the phase space ${\bf C}^1$, and as discussed at
the beginning of this Section an integrable Hamiltonian
is obtained by considering functionals only of Cartan
elements of the Heisenberg-Weyl algebra (e.g. the harmonic
oscillator $a^\dagger a$ {\it or} the free particle $a+a^\dagger$).

In the case of a  spherical or hyperbolic geometry, the transformation to
Darboux coordinates on $(\MG2,\omega)$ is given
by the diffeomorphism $(z,\zbar)\to(v(z,\zbar),
\overline{v}(z,\zbar))$, where
$$v(z,\zbar)={f_1(z)\over\left({|K|\over4}+\,{\rm sign}(K)
f_1(z)\overline{f}_1(\zbar)\right)^{1/2}}$$
maps $\MG2$ onto $D_K^2=\{z\in{\bf C}^1:\,
{\rm sign}(K)z\zbar\leq\,{\rm sign}(K)\}$, the
unit disk $D^2$ in $\IR^2$ for $K>0$ and its
complement $\IR^2\backslash\,{\rm int}(D^2)$ for $K<0$. The observable
(3.15) in Darboux coordinates is therefore
$$H_1^D(z,\zbar)=\Omega_0z\zbar+\left(\overline{\alpha}_0z
+\alpha_0\zbar\right)\left(1-\,{\rm sign}(K)z\zbar\right)^{1/2}
\quad;\quad z\in D_K^2.\eqno(4.2)$$
Equation (4.2) shows that the other general Hamiltonian
systems defined by (3.15) above are just holomorphic copies
of each other, as they all map onto the same Darboux
system defined by (4.2), namely the quasi-displaced harmonic
oscillator; i.e.
$$z\to z+\alpha_0(1-\,{\rm sign}(K)z\zbar)^{1/2}.$$
We therefore consider the quantum problems defined on the phase spaces
$S^2$ and $H^2$ only, and normalize the local coordinates so that now $|K|=1$.

The three independent observables in (3.15)
on the sphere are just the realization of the
SU(2) Lie algebra on $S^2$ by its standard Poisson
structure. Indeed, if we write
$$J_3(z,\zbar)=-j{1-z\zbar\over1+z\zbar}\quad,\quad
J_+(z,\zbar)=2j{\zbar\over1+z\zbar}\quad,\quad J_-(z,\zbar)=2j
{z\over1+z\zbar},\eqno(4.3)$$
then the Poisson algebra of these functions
$$\{J_3,J_\pm\}=\pm J_\pm\quad,\quad\{J_+,J_-\}=2J_3$$
is just SU(2). The general Hamiltonians ${\cal F}(H)$
defined by (3.15) are then functions on the coadjoint
orbit ${\rm SU}(2)/{\rm U}(1)=S^2$
of SU(2), and the corresponding path integral gives
the quantization of spin [16]. The generators
(4.3) are in fact the matrix elements $J_i(z,\zbar)=
(z|J_i|z)/(z|z)$ in the SU(2) coherent states [15]
$$|z)=\e^{zJ_+}|j,-j> { } =\sum_{m=-j}^j\pmatrix
{2j\cr j+m\cr}^{1/2}z^{j+m}|j,m>\quad;\quad z\in{\bf C}^1$$
for the irreducible spin-$j$ representation of SU(2),
$j={1\over2},\,1,\ldots$.

As before, an integrable Hamiltonian is obtained
by considering functionals only of Cartan elements, which
for SU(2) means taking $H=J_3$ above, the height function on $S^2$.
The WKB localization formula (2.4) was computed in
[3,4,17] and the general formula (2.5) for ${\cal F}(H)=H$ in [7],
and shown to give the exact Weyl character formula for SU(2) [16]
$$Z(T|-j,0;S^2,H)={\sin(2j+1){T\over2}\over\sin{T\over2}}.$$
The formula (2.5) was considered in this case for
${\cal F}(H)=H^2$ as well in [7] and shown to give the
exact result
$$Z(T|-j,0;S^2,H^2)=\sum_{m=-j}^j\e^{-iTm^2}=\,{\rm Tr}_{(j)}
\,\e^{-iTJ_3^2}.$$
The path integrals here yield the quantization
of the harmonic oscillator on the sphere (see (4.2)),
and therefore the only integrable quantum system,
up to holomorphic equivalence, we can consider within the
equivariant localization framework on a general
spherical geometry is the harmonic oscillator (on the reduced
phase space $D^2$).

The situation is analogous for the function (3.15)
on the Lobaschevsky plane $H^2$. If we define
$$S_3(z,\zbar)=k{1+z\zbar\over1-z\zbar}\quad,\quad
S_+(z,\zbar)=2k{\zbar\over1-z\zbar}\quad,\quad S_-(z,\zbar)=
2k{z\over1-z\zbar}\eqno(4.4)$$
then the Poisson algebra of these observables is just the
SU(1,1) algebra
$$\{S_3,S_\pm\}=\pm S_\pm\quad,\quad\{S_+,S_-\}
=-2S_3.$$
The Hamiltonians obtained from (3.15) are therefore functions on
the coadjoint orbit ${\rm SU}(1,1)/\,{\rm U}(1)=H^2$ of the non-compact
Lie group SU(1,1) with the generators (4.4)
being matrix elements in the SU(1,1) coherent states
$$|z)=\e^{zS_+}|k;0> { } =\sum_{n=0}^\infty
\pmatrix{2k+n+1\cr n\cr}^{1/2}z^n|k;n>\quad;\quad z\in{\rm int}(D^2)$$
for the discrete representation of SU(1,1) characterized
by $k=1,\,{3\over2},\,2,\,{5\over2},\ldots$ [15,16,18].
Again integrable Hamiltonian systems are obtained by
taking $H=S_3$, which is the height function on $H^2$,
and the corresponding path integral
gives the quantization of the harmonic oscillator
on the open infinite space $H^2$ (and up to holomorphic equivalence
these are the only integrable systems possible
on a general hyperbolic geometry). The WKB approximation for this
coadjoint orbit path integral has been shown to be exact
and gives the usual Weyl character formula for SU(1,1) [17]
$$Z(T|k,0;H^2,H)=2i{\e^{-iT\left(k-{1\over2}\right)}\over
\sin{T\over2}}.$$

\medskip
{\noindent
{\it 4.2. Phase Spaces with One-parameter Isometry Groups}}

In the case where $(\MG2,g)$ isn't maximally symmetric,
we showed in Section 3.4 that it can have local coordinates
defined on it such that its single isometry can be taken
to be translations in a single coordinate, and then
the Hamiltonian, in the coordinates (3.1), is given by the corresponding
transformation function of the {\it other} coordinate, which
is constant along the integral curves of the isometry generator. For
example, if the metric (3.1) is initially
radially symmetric (i.e. $\varphi$ depends only on the product $z\zbar$)
then the desired coordinate transformation is to the usual polar coordinates
$$\eqalign{\theta(x^1,x^2)&=\arctan\left({x^2
\over x^1}\right)\cr r^2(x^1,x^2)&=(x^1)^2+(x^2)^2\cr}\eqno(4.5)$$
in which the Killing vector $V=\partial_\theta$ generates
translations in $\theta$. In the original coordinates
on $\MG2$ the corresponding Hamiltonian
function is nothing but the harmonic oscillator
$H={1\over2}r^2$, as expected from
(4.5) and the general arguments of Section 3.4 above.

Thus one would initially expect that the localization
formulas (2.4) and (2.5) are exact for the harmonic
oscillator with any radially symmetric geometry
on the underlying phase space. This is true of
the WKB formula (2.4), wherein the phase space metric does
not enter directly into its evaluation, but
the more general formula (2.5) is explicitly
metric dependent through the equivariant $\hat{A}$-genus.
In fact, it was shown by Dykstra {\it et al.} [14] that (2.5)
(with ${\cal F}(H)=H$) gives the correct
result for the harmonic oscillator path integral
{\it only} if the phase space metric satisfies the additional constraint
$$\lim_{r\to0}{\partial_rg_{\theta\theta}(r)
\over\sqrt{g(r)}}=-1\eqno(4.6)$$
where $g(r)=\,{\rm det}_{(\mu\nu)}[g_{\mu\nu}(r)]$.
For the metric (3.1) with radial symmetry, this implies that
$$\lim_{r\to0}r{d\over dr}\varphi(r)=0\eqno(4.7)$$
so that the conformal factor $\varphi(r)$ must be an
analytic function of $r$ about $r=0$. This puts a
somewhat strong restriction on the general
form of the metric (3.1) (i.e. on the functional
properties of the conformal factor $\varphi$) and it ensures
that $Z(T)$ is independent of the phase space metric, as it
should be. As discussed in [14], this appears to
be a general problem with the more general
localization formulas (2.5), in that one must
essentially know the quantum theory {\it ab initio} in order
to resolve the ambiguities associated with the
arbitrariness of the metric (3.1). Nevertheless, the
above simple example shows that quite general geometries
can still be used to equivariantly localize the path
integral (1.4) using the formulas (2.4) and (2.5).

This problem can also be seen to be the case for quantum systems which
are not WKB exact. In [14] a similar metric ambiguity
was shown to occur for the one dimensional hydrogen atom
Hamiltonian $H(x^1,x^2)={1\over2}(x^1)^2-1/|x^2|$, a quantum
system whose classical bound state orbits coalesce and so
cannot be WKB localized [23]. They argued that the localization formula
(2.5) with ${\cal F}(H)=H$ is however exact for this quantum
mechanical example. In our context this Hamiltonian would occur, following
the analysis of Section 3.4, if the phase space metric
$g_{\mu\nu}(x^1,x^2)$ in (3.1) is invariant under translations
$\tilde{\chi}^1\to\tilde{\chi}^1+a_0$ of the variable
$$\tilde{\chi}^1(x^1,x^2)=-{1\over\left|{2\over|x^2|}-(x^1)^2
\right|^{3/2}}\left[x^1|x^2|\sqrt{\left|{2\over|x^2|}-(x^1)^2\right|}
+2\arctan\left({x^1\over\sqrt{\left|{2\over|x^2|}-(x^1)^2\right|}}
\right)\right].\eqno(4.8)$$
The coordinate transformation $(x^1,x^2)\to(\tilde{\chi}^1(x^1,x^2),
\tilde{\chi}^2(x^1,x^2))$, where $\tilde{\chi}^1$ is given by (4.8) and
$(\tilde{\chi}^2(x^1,x^2))^2=2/|x^2|-(x^1)^2$, then produces
new coordinates $x'$ in which the metric is independent
of $x'^1$ and the Killing vector has the single
non-vanishing component $V'^1=1$. From (3.18) we see that
under these circumstances we do indeed obtain the
hydrogen atom Hamiltonian $H=-{1\over2}(\tilde{\chi}^2)^2$.
The problem, however, is the existence of a well-defined metric
which is translation invariant in the variable (4.8), but
the results of [14] show that such a metric can be globally defined on
the phase space, and then imposing a further constraint similar
to (4.6) the localization formula (2.5) gives the usual hydrogen
atom partition function [14,23]
$$Z\Bigl(T\Bigm|-{1\over2};\tilde{\chi},H\Bigr)=
\sum_{n=0}^\infty\e^{iT/2n^2}.$$

This example shows that non-trivial quantum systems
can arise from the equivariant localization constraints,
but only for phase space geometries which have complicated
and unusual symmetries (such as translations in (4.8) above).
Thus aside from the above noted problem of resolving the metric
ambiguity in the localization formulas, there is
further the problem as to whether or
not a two dimensional geometry can in fact possess
the required symmetry. Of course we do not expect that
any two dimensional quantum mechanical Hamiltonian will
have an exactly solvable path integral, and in the
present point of view the cases where the path integral
fails to be effectively computable will
be the cases where a required symmetry of
the phase space geometry does not lead to a globally well-defined
metric tensor. For example, in [14] it was argued
that, with the exception of the harmonic oscillator,
equivariant localization fails for all one dimensional
quantum mechanical Hamiltonians with static potentials
which are bounded below, due to the non-existence of a
single-valued metric satisfying the Killing equation
(2.3) in this case.

We can alternatively formulate the quantum theory of
systems with non-homogeneous phase spaces within
the framework of Section 4.1 by using a formalism for
coherent states associated with non-transitive group actions [19].
Let us consider the metric (3.1) in the coordinates
$x'$ described in Section 3.4, which we label
by $z=x'^2\e^{ix'^1}$. Let $f(z\zbar)$ be an analytic
solution of the ordinary differential equation
$${d\over d(z\zbar)}z\zbar{d\over d(z\zbar)}
\log f(z\zbar)={1\over2}\e^{\varphi(z\zbar)}.\eqno(4.9)$$
The symplectic 2-form can be chosen to be the volume form of $(\MG2,g)$,
$$\omega=i{d\over d(z\zbar)}z\zbar{d\over d(z\zbar)}
\log f(z\zbar)dz\wedge d\zbar,\eqno(4.10)$$
so that the canonical 1-form is
$$\theta={i\over2}{d\over d(z\zbar)}\log f(z\zbar)\left(
\zbar dz-zd\zbar\right).\eqno(4.11)$$

Let $N_\varphi$, $0<N_\varphi\leq\infty$, be the integer such that
the function $f(z\zbar)$ has the Taylor expansion
$$f(z\zbar)=\sum_{n=0}^{N_\varphi}(z\zbar)^nf_n,$$
and let $\rho(z\zbar)$ be an integrable function whose moments are
$$\int_0^Pd(z\zbar)\,(z\zbar)^n \rho(z\zbar)={1\over f_n}\quad,
\quad0\leq n\leq N_\varphi,$$
where $P$ is a real number with $0<P\leq\infty$. Let $a^\dagger$ and
$a$ be bosonic creation and annihilation operators on
some representation space of the isometry group,
and let $|n>$ be the complete system of eigenstates of the
corresponding number operator, $a^\dagger a|n> { } =n|n>$.
The desired coherent states are defined as
$$|z)=\sum_{n=0}^{N_\varphi}\sqrt{f_n}z^n|n>.\eqno(4.12)$$
These states have the normalization $(z|z)=f(z\zbar)$ and
satisfy the completeness relation
$$\int_{\MG2}d\mu(z,\zbar)\,{|z)(z|\over(z|z)}={\bf 1}\eqno(4.13)$$
in the U(1)-invariant measure
$$d\mu(z,\zbar)={i\over2\pi}f(z\zbar)\rho(z\zbar)
\Theta(P-z\zbar)dz\wedge d\zbar.$$
Notice that for $f(z\zbar)=\e^{z\zbar}$, $(1+z\zbar)^{2j}$ and
$(1-z\zbar)^{-2k}$, (4.12) reduces to, respectively, the
Heisenberg-Weyl group, spin-$j$ SU(2) and level-$k$ SU(1,1)
coherent states in Section 4.1 above, as anticipated from (4.9).
The isometry group acts on the states (4.12) as
$V_\tau|z)=|\e^{i\tau}z)\, ; \, V_\tau\in{\cal I}(\MG2,g),
\tau\in\IR^1$, and the metric tensor (3.1) and symplectic 1-form
(4.11) can be expressed in the standard coherent state forms [15]
$$g=4\left[{\|d|z)\|\over\sqrt{(z|z)}}\otimes{\|d|z)\|\over
\sqrt{(z|z)}}-{(z|d|z)\over(z|z)}\otimes{\overline{(z|d|z)}\over(z|z)}\right]$$
$$\theta={i(z|d|z)\over(z|z)}.$$

As before, we consider the matrix elements $H(z,\zbar)=
(z|{\cal H}|z)/(z|z)$ of some operator $\cal H$ on the
underlying Hilbert space, and then using (4.11) and (4.13) the standard
coherent state path integral may be constructed as [15]
$$\eqalign{Z(T|a_0;f,{\cal F}(H))=&\int_{L\MG2}\prod_t
d\mu(z(t),\zbar(t))\cr&\qquad\times\exp\biggl[i\int_0^T
dt\,\biggl({1\over2}{d\over d(z\zbar)}\log f(z\zbar)
\left(z\dot{\zbar}-\zbar\dot{z}\right)-{\cal F}(H)
\biggr)\biggr].\cr}\eqno(4.14)$$
The observable $H(z,\zbar)$ appearing in (4.14) can now be
found as before from the equivariant localization
constraints and is given in terms of the phase space metric as
$$H(z,\zbar)=a_0z\zbar{d\over d(z\zbar)}\log f(z\zbar)
+C_0=a_0{(z|a^\dagger a|z)\over(z|z)}+C_0\eqno(4.15)$$
where the function $f(z\zbar)$ is related to the metric (3.1)
by equation (4.9). For the maximally symmetric cases
of Section 4.1 above (4.15)
reduces to the observables $H$ given there (as does the
coherent state path integral (4.14) for these cases)
\footnote{$^4$} {\tenpoint For the cases of the
Heisenberg-Weyl, SU(2) and SU(1,1) group actions on
$\MG2$, the weight functions are $\rho(z\zbar)=
\e^{-z\zbar}$ ($P=\infty$), $(2j+1)(1+z\zbar)^{-2(j+1)}$
($P=\infty$) and $(2k-1)(1-z\zbar)^{2(k-1)}$
($P=1$), respectively.}. Thus the formula
(4.15) is a general formula valid for {\it any} geometry on
the underlying phase space, be it maximally symmetric or otherwise.
This is not surprising, since as remarked in Section 4.1 the
Hamiltonian functions obtained for the case of maximal symmetry
are just displaced harmonic oscillators, and the oscillator
Hamiltonians arise from the rotation generators of the
isometry groups, i.e. translations in ${\rm arg}(z)=x'^1$,
and so the coordinate system used in Section 4.1 coincides with
that defined in Section 3.4 and used in (4.15) (this also
agrees with what one expects from integrability arguments). In fact, the
expression (4.15) shows explicitly that the function $H$ is really just
a harmonic oscillator Hamiltonian on some general geometry.

The main difference in the present context between the maximally symmetric and
non-homogeneous cases lies in the path integral (4.14) itself:
In the former case the measure $d\mu(z,\zbar)$ which must be
used in the Feynman measure in (4.14) coincides with the
volume form (4.10), because if the isometry group acts transitively
on $(\MG2,g)$ then there is a unique left-invariant
measure (i.e. a unique solution to equation (3.8))
and so $d\mu=\omega$ yields the standard Liouville measure
on the loop space. In the latter case $d\mu\neq\omega$
in general (i.e. (4.14) is not necessarily in the form (1.4),
but the standard localization formulas still hold with the
obvious replacements corresponding to the change of integration measure).
Of course the standard Liouville path integral measure can
be used if instead one follows the prescription of Section 3.4.
It is essentially the non-uniqueness of a U(1)-invariant
symplectic 2-form in the case of non-transitive isometry group
actions which leads to numerous possibilities for the
Hamiltonians on such geometries. If one consistently makes
the ``natural" choice of the volume form (4.10), then indeed
the only admissible Hamiltonian functions $H$ are generalized
harmonic oscillators.

\bigskip

{\noindent
{\bf 5. Summary and Discussion}}
\medskip

In this Paper we have discussed what Hamiltonian systems
with two dimensional simply connected phase spaces can
have equivariantly localized Feynman path integrals.
We derived general formulas for the Hamiltonian functions
in terms of the underlying phase space geometry, which
shows how equivariant localization of path integrals is
explicitly metric dependent and provides a
picture of how the quantum geometry affects the exact
solvability of some two dimensional quantum mechanical systems.
This is interesting in that the underlying quantum
theory is always {\it ab initio} metric-independent
so that an analysis such as the one
presented here probes into how the classical phase
space geometry is modified by quantum effects and the role
geometry plays towards the understanding of quantum integrability. For
instance, we showed that the classical trajectories of
a harmonic oscillator must be embedded, at the quantum
level, into a rotationally invariant geometry. For
more complicated systems the quantum geometries
are less familiar and endow the phase space with unusual
Riemannian structures.

We showed that all Hamiltonians so obtained are essentially
harmonic oscillators, with modification terms to take into account
the possible non-trivial geometry (3.1) of the phase space
(see (4.15)). These observables arise naturally in coherent
state formalisms corresponding to the Poisson Lie group actions
of the appropriate isometry groups on the phase space, and the
fact that they always correspond rather trivially to harmonic
oscillator Hamiltonians is equivalent to the
original constraint that they generate a circle action on the
phase space. Non-trivial systems (such as the one dimensional
hydrogen atom for which the semi-classical approximation is unsuitable)
arise only under somewhat ad-hoc restrictions on the symmetries
the given geometry can possess (see (4.8)) and there seems
to be no general way to obtain the set of all Hamiltonians
which arise from a general non-homogeneous geometry. Although
we have obtained a formal prescription in Section 3.4 which
in principle allows one to obtain such systems, in practice
introducing such a definite geometry into the problem is
quite non-trivial. For these latter non-homogeneous cases
there is also the problem that the quantum theory must be
essentially known before hand in order to resolve the
ambiguities associated with the single degree of freedom of
the metric (3.1). Thus the localization formulas, although
being nice formal insights into quantum integrability
and its geometric nature, are only practically applicable
to rather trivial systems. Our results here also impose restrictions
the class of topological quantum field theories and
supersymmetric models which arise from these geometric
localization principles [3--5].

Of course our analysis and comments above are all made under the
rather severe topological restriction $\pi_1(\MG2)=0$ of
simple connectivity of the phase space. We expect that
more non-trivial systems can arise for more complicated
topologies, such as the 2-torus $T^2=S^1\times S^1$,
which has fundamental group $\pi_1(T^2)={\bf Z}\oplus
{\bf Z}$. For example, it is known
that the Duistermaat-Heckman formula (1.2) fails for the
Hamiltonian system whose phase space is the
torus and whose Hamiltonian function is the height
function on $T^2$ [4].

Notice further that the analysis presented in this Paper
carries through rather nicely only in two dimensions.
For higher dimensional symplectic manifolds
$(\Gamit,\omega)$ ($n>1$), the possibilities
are far more numerous. First of all, the phase space metric
will have more than one degree of freedom, and a general
expression of the form (3.1) is not possible. Thus a complete
geometric classification as above is not possible. Secondly,
the isometry group ${\cal I}(\Gamit,g)$ can now have up to
$n(2n+1)$ generators [10], and the total number of possible dimensions
of the isometry group will in general be more than 2 (the case
of isometries of two dimensional Riemannian manifolds is quite
unique in its properties as compared to
higher dimensions [10]). There are also the
additional cases where the manifold itself isn't maximally
symmetric but contains a smaller, maximally symmetric
subspace. Thirdly, even
for the standard maximally symmetric spaces in $2n$ dimensions,
whose Killing vectors can be readily constructed [10],
the Hamiltonian equations (1.3) are more difficult to solve
because the symplectic 2-form is no longer a top-form
on the manifold. We therefore expect that our analysis here
is rather difficult to generalize to arbitrary dimensional
symplectic manifolds. Nonetheless, the two dimensional
analysis presented here gives an indication of the range of
applicability of the localization formulas in general.

\bigskip

\centerline{\bf Acknowledgements}
\medskip

We wish to thank M. Bergeron, E. Langmann, O. Tirkkonen
and W. Unruh for helpful discussions. This work was supported
in part by the Natural Sciences and Engineering Research
Council of Canada.

\vfill\eject

\singlespace

\centerline{\bf References}

\bigskip

\item{[1]} L. S. Schulman, {\it Techniques and Applications
of Path Integration}, John Wiley and Sons (New York) (1981).
\line{\hfill}
\item{[2]} M. Blau, E. Keski-Vakkuri and A. J. Niemi,
Phys. Lett. {\bf B246} (1990), 92.
\line{\hfill}
\item{[3]} A. J. Niemi and P. Pasanen, Phys. Lett.
{\bf B253} (1991), 349.
\line{\hfill}
\item{[4]} E. Keski-Vakkuri, A. J. Niemi, G. W. Semenoff
and O. Tirkkonen, Phys. Rev. {\bf D44} (1991), 3899.
\line{\hfill}
\item{[5]} A. Hietam\"aki, A. Yu. Morozov, A. J. Niemi and
K. Palo, Phys. Lett. {\bf B263} (1991), 417; Nucl. Phys.
{\bf B377} (1992), 295; A. Yu. Morozov, A. J. Niemi and
K. Palo, Phys. Lett. {\bf B271} (1991), 365; A. Hietam\"aki
and A. J. Niemi, ``Index Theorems and Loop Space Geometry",
CERN Preprint CERN-TH 6471 (1992); A. P. Nersessian,
JETP Lett. {\bf 58} (1993), 64; ``Equivariant Localization:
BV-geometry and Supersymmetric Dynamics", Joint Institute
for Nuclear Research Preprint (1993); K. Palo, ``Symplectic
Geometry of Supersymmetry and Nonlinear Sigma Model",
University of Uppsala Preprint UU-ITP 30/93 (1993).
\line{\hfill}
\item{[6]} A. J. Niemi and O. Tirkkonen, Phys. Lett.
{\bf B293} (1992), 339.
\line{\hfill}
\item{[7]} A. J. Niemi and O. Tirkkonen, ``On Exact Evaluation
of Path Integrals", University of Uppsala Preprint UU-ITP 3/93
(1993).
\line{\hfill}
\item{[8]} A. J. Niemi and K. Palo, Mod. Phys. Lett.
{\bf A8} (1993), 2311.
\line{\hfill}
\item{[9]} E. Witten, J. Geom. Phys. {\bf 9} (1992), 303.
\line{\hfill}
\item{[10]} L. P. Eisenhart, {\it Riemannian Geometry}, Princeton
University Press (Princeton) (1949); {\it Continuous Groups of
Transformations}, Dover Publications (New York) (1961);
S. Helgason, {\it Differential Geometry, Lie Groups and
Symmetric Spaces}, Academic Press (New York) (1978).
\line{\hfill}
\item{[11]} J. J. Duistermaat and G. J. Heckman, Inv.
Math. {\bf 69} (1982), 259.
\line{\hfill}
\item{[12]} N. Berline and M. Vergne, Duke Math. J. {\bf 50}
(1983), 539; V. Guillemin and S. Sternberg, {\it Symplectic
Techniques in Physics}, Cambridge University Press (Cambridge)
(1984); M. F. Atiyah and R. Bott, Topology {\bf 23} (1984), 1;
M. F. Atiyah, Asterisque {\bf 131} (1985), 43; J.-M. Bismut,
Comm. Math. Phys. {\bf 98} (1985), 213.
\line{\hfill}
\item{[13]} J. A. Minahan and A. P. Polychronakos, ``Classical
Solutions for Two Dimensional QCD on the Sphere", CERN Preprint
CERN-TH-7016/93 (1993).
\line{\hfill}
\item{[14]} H. M. Dykstra, J. D. Lykken and E. J. Raiten,
Phys. Lett. {\bf B302} (1993), 223.
\line{\hfill}
\item{[15]} A. M. Perelomov, {\it Generalized Coherent States
and Their Applications}, Springer-Verlag (Berlin) (1986);
L. Faddeev, in {\it Methods in Field Theory}, Proceedings of
Les Houches Summer School, Les Houches, France, 1975, eds.
R. Balian and J. Zinn-Justin, Les Houches Summer School
Proceedings Vol. XXVIII, North-Holland (Amsterdam) (1976);
J. R. Klauder and B.-S. Skagerstam, {\it Coherent States},
World Scientific (Singapore) (1985).
\line{\hfill}
\item{[16]} A. Alekseev, L. Faddeev and S. Shatashvili,
J. Geom. Phys. {\bf 1} (1989), 3; A. Alekseev and S. Shatashvili,
Nucl. Phys. {\bf B323} (1989), 719; H. B. Nielsen and
D. Rohrlich, Nucl. Phys. {\bf B299} (1988), 471; M. Stone,
Nucl. Phys. {\bf B314} (1989), 557; O. Alvarez, I. M. Singer
and P. Windey, Nucl. Phys. {\bf B337} (1990), 467; M. Blau,
Int. J. Mod. Phys. {\bf A6} (1991), 365.
\line{\hfill}
\item{[17]} S. G. Rajeev, S. Kalyana Rama and Siddhartha Sen,
``Symplectic Manifolds, Coherent States and Semiclassical
Approximation", University of Rochester Preprints UR-1324,
ER 40685-774 (1993).
\line{\hfill}
\item{[18]} A. M. Perelomov, Comm. Math. Phys. {\bf 26} (1972),
222; F. A. Berezin, Comm. Math. Phys. {\bf 40} (1975), 153;
{\it The Method of Second Quantization}, Nauka (Moscow) (1986).
\line{\hfill}
\item{[19]} J. R. Klauder, Mod. Phys. Lett. {\bf A8} (1993), 1735.
\line{\hfill}
\item{[20]} N. Berline, E. Getzler and M. Vergne, {\it Heat
Kernels and Dirac Operators}, Springer-Verlag (Berlin) (1991).
\line{\hfill}
\item{[21]} V. I. Arnol'd and S. P. Novikov, {\it Dynamical
Systems IV}, Springer-Verlag (Berlin) (1990).
\line{\hfill}
\item{[22]} A. Das, {\it Integrable Models}, World Scientific
(Singapore) (1989).
\line{\hfill}
\item{[23]} R. Loudon, Am. J. Phys. {\bf 27} (1959), 649.

\end